\input{aipcheck}

\documentclass[
    ,final            % use final for the camera ready runs
  ]
  {aipproc}

\layoutstyle{6x9}

\usepackage{graphicx,color,here,epsfig}

\begin{document}

\title{Double spin azimuthal asymmetries $A_{LT}$ and $A_{LL}$ in semi-inclusive DIS }


\author{Aram Kotzinian}{
  address={{\it Yerevan Physics Institute, 375036 Yerevan, Armenia} \\
  {\it Dipartimento di Fisica Generale, Universit\`a di Torino, and \\
INFN, Sezione di Torino, Via P. Giuria 1, I-10125 Torino, Italy}\\
{\it $^2$\it JINR, 141980 Dubna, Russia}} }

\begin{abstract}Within the LO QCD parton model of SIDIS,
$\ell \, N \to \ell \, h \, X$, with unintegrated quark distribution
and fragmentation functions, we study the transverse momentum and
azimuthal dependencies of the double spin asymmetries $A_{LT}$ and
$A_{LL}$. For later we include ${\cal O}(k_{\perp}/Q)$ kinematic
corrections, which induce an azimuthal modulation of the asymmetry,
analogous to the Cahn effect in unpolarized SIDIS. We show that a
study of these asymmetries allows to extract the transverse momentum
dependence of the unintegrated helicity distribution function
$g_{1L}^q(x,k_\perp)$ and $g_{1T}^q(x,k_\perp)$.

This report is based on research published in~\cite{kpp,aekp}, where
predictions are given for ongoing COMPASS, HERMES and JLab
experiments.
\end{abstract}

\maketitle

%%%%%%%%%%%%%%%%%%%%%%%%%%%%%%%%%%%%%%%%%%%%
%% MAINMATTER
%%%%%%%%%%%%%%%%%%%%%%%%%%%%%%%%%%%%%%%%%%%%

%\section{\label{sec:intro}Introduction}

Following Ref.~\cite{ko}, we consider the polarized SIDIS processes
at twist-two in the parton model with transverse momentum dependent
distribution and fragmentation functions (TMD DFs and FFs), taking
into account ${\cal O}(k_{\perp}/Q)$ kinematical correction for
$A_{LL}$ asymmetries (so called Cahn effect~\cite{cahn}).

The cross section for polarized SIDIS can be written as:
\begin{equation}\label{dsll}
\frac{d^5\sigma^{pol}}{dx \, dy \, dz \, d^2P_{hT}} = \frac{2
\alpha^2}{x y^2 s}\, \left\{ {\cal H}_{f_1} + \lambda \, (S_L {\cal
H}_{g_{1L}} + S_T{\cal H}_{g_{1T}})+... \right\}. \label{sig}
\end{equation}
In the case of longitudinally polarized target, where longitudinal
(according to the laboratory setup) refers to the initial lepton
direction, one get a transverse -- with respect to the $\gamma^*$
direction -- spin component:
\begin{equation}\label{sintg}
    S_T = S\sin\theta_\gamma,\>\>
\sin\theta_\gamma = \sqrt{\frac{4M^2x^2}{Q^2+4M^2x^2} \left( 1-y -
\frac{M^2x^2y^2}{Q^2} \right)} \simeq \frac{2Mx\sqrt{1-y}}{Q}\;\cdot
\end{equation}
This component gives contributions of order $M/Q$.

We assumes a simple factorized and gaussian behavior of the involved
TMD PDFs and FFs
\begin{eqnarray}
f^q_1(x,k_\perp) &=& f^q_1(x) \, \frac{1}{\pi \mu_0^2}\, \exp\left(
-\frac{k_\perp^2}{\mu_0^2} \right),\> D_q^h(z, p_\perp)= D_q^h(z) \,
\frac{1}{\pi \mu_D^2}\,
\exp\left( -\frac{p_\perp^2}{\mu_D^2} \right)\label{dfffg1},\\
g^{q\perp}_{1T}(x,k_\perp)&=& g^{q}_{1T}(x) \, \frac{1}{\pi
\mu_1^2}\, \exp\left( -\frac{k_\perp^2}{\mu_1^2} \right),\>
g^q_{1L}(x,k_\perp)= g^q_{1}(x) \, \frac{1}{\pi \mu_2^2}\,
\exp\left( -\frac{k_\perp^2}{\mu_2^2} \right)\label{dfffg2}.
\end{eqnarray}
Following Ref. \cite{anskpr} we use $\mu_0^2 = 0.25 \> ({\rm
GeV}/c)^2, \mu_D^2 = 0.20 \> ({\rm GeV}/c)^2$ while we consider
$\mu_1^2$ and $\mu_2^2$ as free parameters.

For the longitudinally polarized target we consider the $P_{hT}$
dependence of the double longitudinal spin asymmetry
\begin{equation}\label{asym}
A_{LL}(x,y,z,P_{hT}) = \frac{\int_0^{2\pi} \, d\phi_h
\, [ \, d\sigma^{\begin{array}{c}\hspace*{-0.1cm}\to\vspace*{-0.25cm}\\
\hspace*{-0.1cm}\Leftarrow\end{array}}
- d\sigma^{\begin{array}{c}\hspace*{-0.1cm}\to\vspace*{-0.25cm}\\
\hspace*{-0.1cm}\Rightarrow\end{array}} ]} {\lambda S_L
\int_0^{2\pi} \, d\phi_h \, [ \,
d\sigma^{\begin{array}{c}\hspace*{-0.1cm}\to\vspace*{-0.25cm}\\
\hspace*{-0.1cm}\Leftarrow\end{array}}
+ d\sigma^{\begin{array}{c}\hspace*{-0.1cm}\to\vspace*{-0.25cm}\\
\hspace*{-0.1cm}\Rightarrow\end{array}} ]} \>,
\end{equation}
and the $\cos\phi_h$ weighted asymmetry, defined as
\begin{equation}\label{wasym}
A_{LL}^{\cos\phi_h}(x,y,z,P_{hT}) = \frac{2\int_0^{2\pi} \, d\phi_h
\, [ \, d\sigma^{\begin{array}{c}\hspace*{-0.1cm}\to\vspace*{-0.25cm}\\
\hspace*{-0.1cm}\Leftarrow\end{array}}
- d\sigma^{\begin{array}{c}\hspace*{-0.1cm}\to\vspace*{-0.25cm}\\
\hspace*{-0.1cm}\Rightarrow\end{array}} ]\cos\phi_h} {\lambda S_L
\int_0^{2\pi} \, d\phi_h \, [ \,
d\sigma^{\begin{array}{c}\hspace*{-0.1cm}\to\vspace*{-0.25cm}\\
\hspace*{-0.1cm}\Leftarrow\end{array}}
+ d\sigma^{\begin{array}{c}\hspace*{-0.1cm}\to\vspace*{-0.25cm}\\
\hspace*{-0.1cm}\Rightarrow\end{array}} ]} \>\cdot
\end{equation}
Similarly, we define the asymmetry for transversely polarized target
\begin{equation}
A_{LT}^{\cos(\phi_h-\phi_S)}(x,y,z,P_{hT})
=\frac{\int_0^{2\pi}\,d(\phi_h-\phi_S)\,(d\sigma^{\begin{array}{c}\hspace*{-0.1cm}\to\vspace*{-0.25cm}\\
\hspace*{-0.1cm}\uparrow\end{array}}-d\sigma^{\begin{array}{c}\hspace*{-0.1cm}\to\vspace*{-0.25cm}\\
\hspace*{-0.1cm}\downarrow\end{array}})\cos(\phi_h-\phi_S)}
{\lambda S_T \int_0^{2\pi}\,d(\phi_h-\phi_S)\,(d\sigma^{\begin{array}{c}\hspace*{-0.1cm}\to\vspace*{-0.25cm}\\
\hspace*{-0.1cm}\uparrow\end{array}}+d\sigma^{\begin{array}{c}\hspace*{-0.1cm}\to\vspace*{-0.25cm}\\
\hspace*{-0.1cm}\downarrow\end{array}})}, \label{aweight}
\end{equation}
and also asymmetry weighted with $\mathbf{S}_T\cdot
\mathbf{P}_{hT}/M$ = $(\vert\mathbf{
P}_{hT}\vert/M)\cos(\phi_h-\phi_S)$~\cite{km}
\begin{equation}
A_{LT}^{\vert\mathbf{ P}_{hT}\vert/M)\cos(\phi_h-\phi_S)}(x,y,z)
=\frac{\int\,d^{\,2}P_{hT}\,(d\sigma^{\begin{array}{c}\hspace*{-0.1cm}\to\vspace*{-0.25cm}\\
\hspace*{-0.1cm}\uparrow\end{array}}-d\sigma^{\begin{array}{c}\hspace*{-0.1cm}\to\vspace*{-0.25cm}\\
\hspace*{-0.1cm}\downarrow\end{array}})\vert\mathbf{
P}_{hT}\vert/M)\cos(\phi_h-\phi_S)}
{\lambda S_T \int\,d^{\,2}P_{hT}\,(d\sigma^{\begin{array}{c}\hspace*{-0.1cm}\to\vspace*{-0.25cm}\\
\hspace*{-0.1cm}\uparrow\end{array}}+d\sigma^{\begin{array}{c}\hspace*{-0.1cm}\to\vspace*{-0.25cm}\\
\hspace*{-0.1cm}\downarrow\end{array}})}. \label{aweight}
\end{equation}

\begin{figure}[h]
  \hfill
 \hspace{-1.cm}
  \begin{minipage}[t]{.45\textwidth}
    \begin{center}
      \epsfig{file=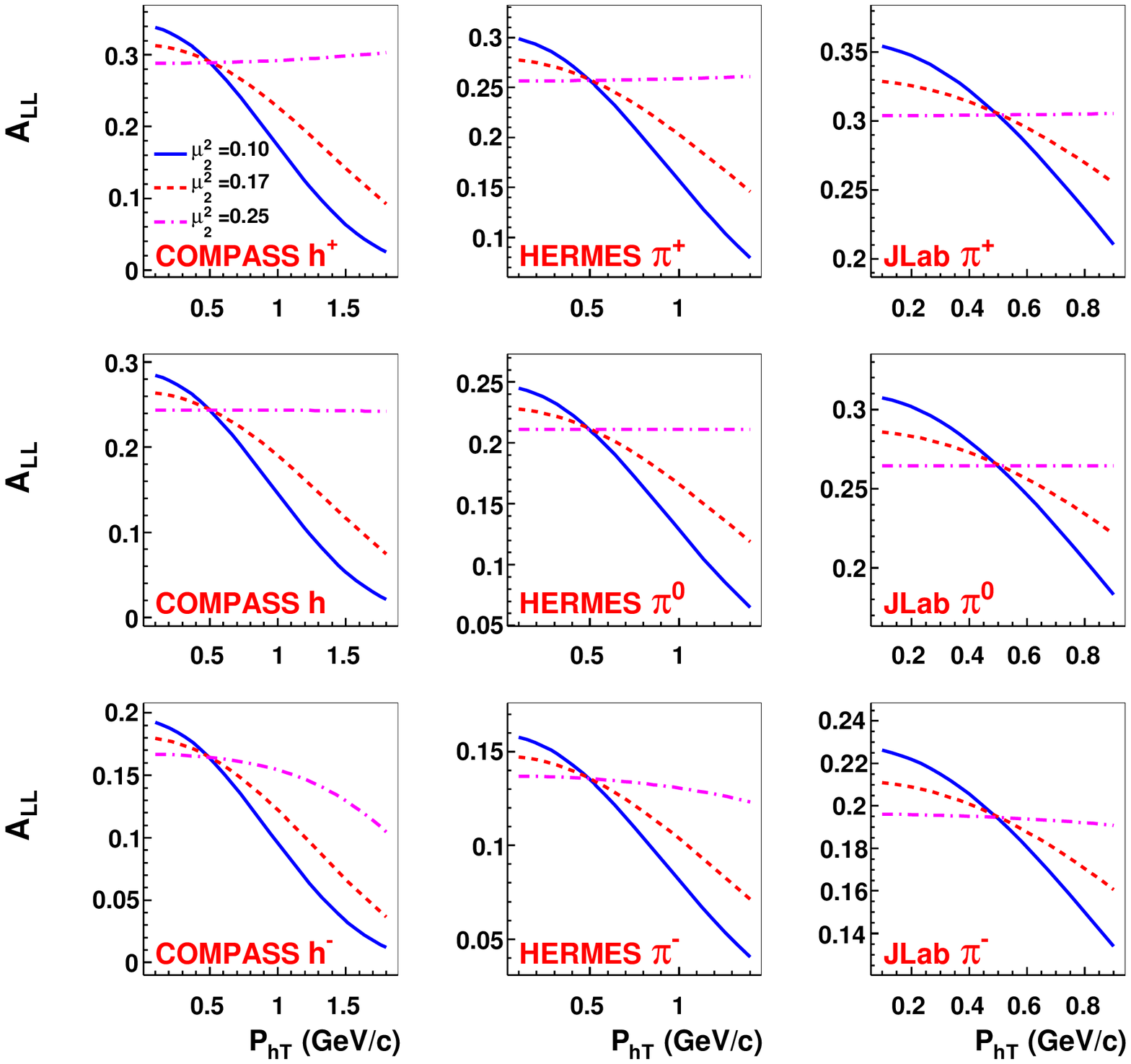, scale=0.38}
      \caption{}
      \label{}
    \end{center}
  \end{minipage}
  \hfill
  \hspace{1.cm}
  \begin{minipage}[t]{.45\textwidth}
    \begin{center}
      \epsfig{file=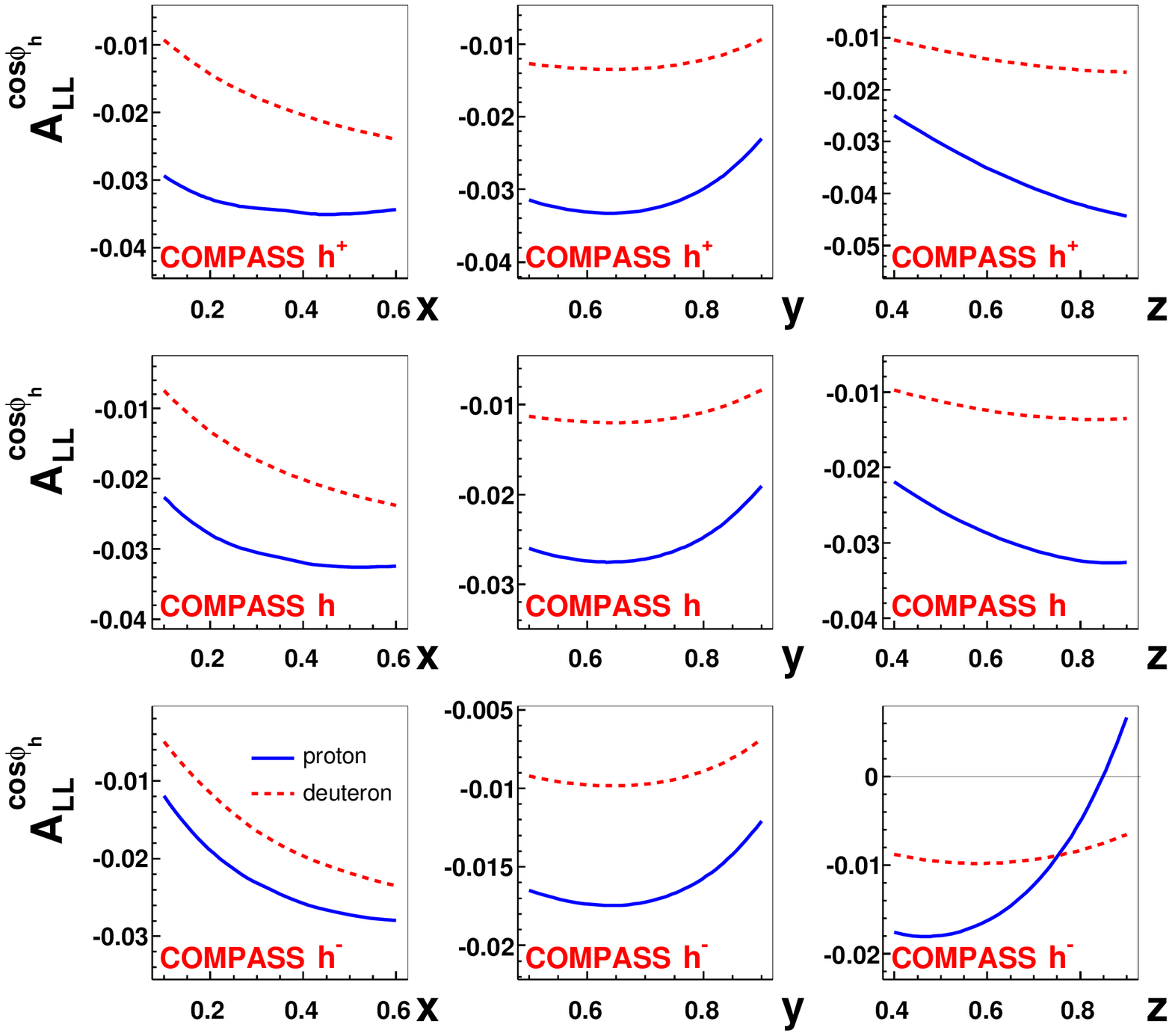, scale=0.38}
      \caption{Left panel: Predicted dependence of
$A_{LL}$ on $P_{hT}$, for scattering off a proton target, with
different choices of $\mu_2^2$: 0.1 (GeV/c)$^2$ -- continuous, 0.17
(GeV/c)$^2$ -- dashed and 0.25 (GeV/c)$^2$ -- dot-dashed lines.
Right panel: predicted dependence of $A_{LL}^{\cos\phi_h}$ on $x$,
$y$ and $z$, for proton -- continuous and deuteron -- dashed line
targets, for COMPASS.}
      \label{fig:all1}
    \end{center}
  \end{minipage}
  \hfill
\end{figure}

As examples of our results we present here some plots from
\cite{kpp, aekp}. According to the range covered by the setups of
the experiments we use the following cuts which are aimed to enhance
asymmetries:
\begin{itemize}
\item {COMPASS: positive ($h^+$), all ($h$) and negative ($h^-$) hadron
production, $Q^2 > 1.0$ (GeV/c)$^2$, $W^2 > 25$ GeV$^2$, $0.1 < x <
0.6$, $0.5 < y < 0.9$ and  $0.4 < z <0.9 $}
\item {HERMES: $\pi^+$, $\pi^0$ and $\pi^-$ production,
$Q^2 > 1.0$ (GeV/c)$^2$, $W^2 > 10$ GeV$^2$, $0.1 < x < 0.6$, $0.45
< y < 0.85$ and  $0.4 < z <0.7 $}
\item {JLab at 6 GeV: $\pi^+$, $\pi^0$ and $\pi^-$ production,
$Q^2 > 1.0$ (GeV/c)$^2$, $W^2 > 4$ GeV$^2$, $0.2 < x < 0.6$, $0.4 <
y < 0.85$ and  $0.4 < z <0.7 $.}
\end{itemize}
Concerning the usual integrated distribution and fragmentation
functions we use the LO GRV98~\cite{grv} unpolarized and the
corresponding GRSV2000~\cite{grsv} polarized (standard scenario)
DFs, and Kretzer~\cite{kretzer} FFs.

\begin{figure}[h]
  \hfill
\hspace{-1.cm}
  \begin{minipage}[t]{.45\textwidth}
    \begin{center}
%    \vspace{1cm}
      \epsfig{file=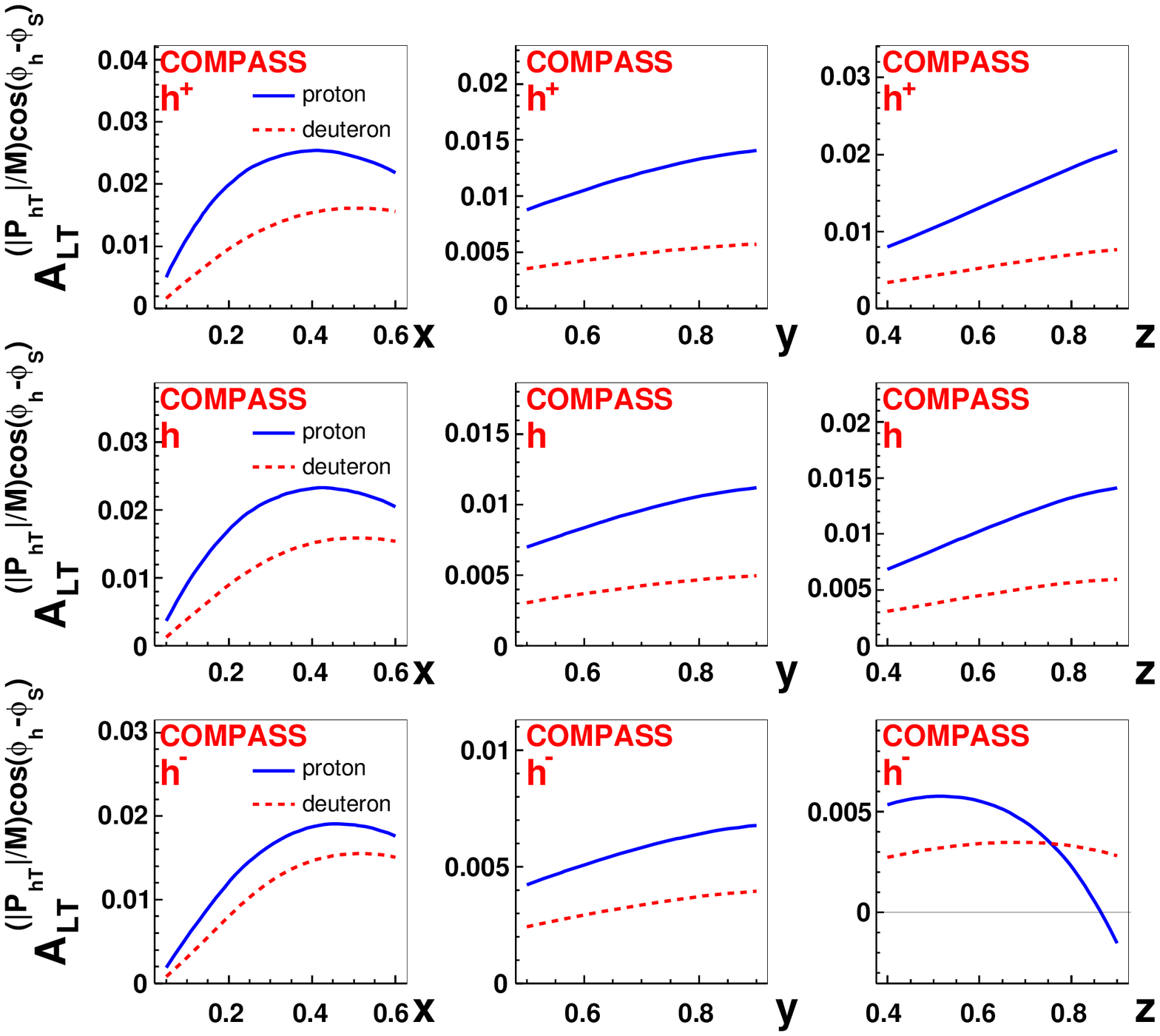, scale=0.38}
      \caption{}
      \label{}
    \end{center}
  \end{minipage}
  \hfill
\hspace{1.cm}
  \begin{minipage}[t]{.45\textwidth}
    \begin{center}
%  \vspace{0.6cm}
      \epsfig{file=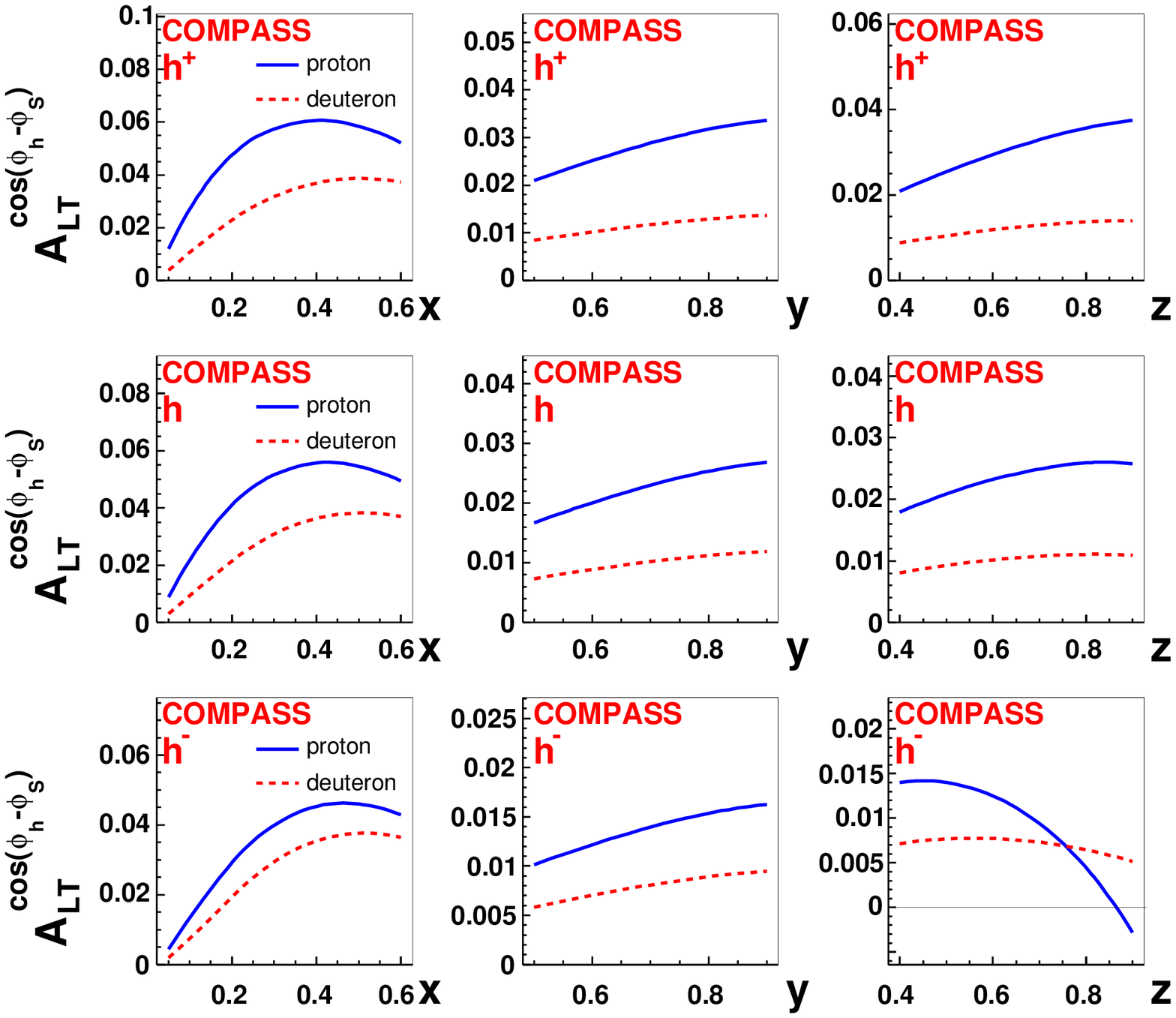, scale=0.38}
      \caption{Left panel: Predicted dependence of $A_{LT}^{(\vert
\mathbf{P}_{hT}\vert/M)\cos(\phi_h-\phi_S)}$ on $x$, $y$ and $z$ for
production of positive ($h^+$), all charged ($h$) and negative
($h^-$) hadrons at COMPASS for SIDIS on transversely polarized
proton -- solid and deuteron -- dashed line targets. Right panel:
predicted dependence of $A_{LT}^{\cos(\phi_h-\phi_S)}$ on $x$, $y$
and $z$ with $\vert \mathbf{P}_{hT, min} \vert = 0.5$ GeV/c for
proton -- continuous and deuteron -- dashed line targets, for
COMPASS.}
      \label{fig:alt1}
    \end{center}
  \end{minipage}
  \hfill
\end{figure}

In the left panel of Fig. \ref{fig:all1} the $P_{hT}$-dependence of
$A_{LL}$ asymmetries are presented. Notice that they are
leading-twist quantities, not suppressed by any inverse power of
$Q$. Although our numerical estimates are based on the gaussian
factorization ansatz, Eqs. \ref{dfffg1}, \ref{dfffg2}, we expect
them to have a more general interpretation and information content.
In the right panel of Fig.~\ref{fig:all1} we present the $x$-, $y$-
and $z$-dependencies of $A_{LL}^{\cos\phi_h}$ integrated over
$P_{hT}$ with $P_{hT, min} = 0.5$ GeV/c and $\mu_1^2=$0.15
(GeV/c)$^2$, $\mu_2^2=$0.25 (GeV/c)$^2$ for COMPASS.

To calculate the double spin asymmetry for transversely polarized
target the DF $g_{1T}^q(x)$ is needed. For the first
$k_{\perp}$-momentum  we have
%defined as $g^{q\perp}_{1T}(x,k_T^2)$
%
\begin{equation}
g_{1T}^{q\,(1)}(x) \equiv \int d^{\,2}k_\perp \,
\frac{k_\perp^2}{2M^2} \> g^{q\perp}_{1T}(x,k_T^2) =
\frac{\mu_1^2}{2M^2} \, g_{1T}^q(x) \label{g1t1} \>.
\end{equation}
Using the Lorentz invariance relations \cite{mt} and
Wandzura-Wilczek approximation \cite{ww} the following relation
\cite{km} has been derived
\begin{equation}
g_{1T}^{q(1)}(x) \simeq x\int_x^1 dx'\,\frac{g^q_1(x')}{x'} \>,
\label{g11tww}
\end{equation}
which allows to express $g_{1T}^{q}(x)$ through the well known
integrated helicity distributions.

In the left panel of Fig. \ref{fig:alt1} the predictions for
$A_{LT}^{(\vert \mathbf{P}_{hT}\vert/M)\cos(\phi_h-\phi_S)}$
asymmetry dependence on $x$, $y$ and $z$ are shown for production of
positive ($h^+$), all charged ($h$) and negative ($h^-$) hadrons at
COMPASS. In the right panel of Fig.~\ref{fig:alt1} we present the
predicted dependence of $A_{LT}^{\cos(\phi_h-\phi_S)}$ on $x$, $y$
and $z$ with $\vert \mathbf{P}_{hT, min} \vert = 0.5$ GeV/c for
proton -- continuous and deuteron -- dashed line targets, for
COMPASS.

The measurement of discussed asymmetries will allow
\begin{itemize}
\item to extract the $k_T$-dependence of TMD DFs,
%$g^{q\perp}_{1T}(x,k_T^2)$ and $g^q_{1L}(x,k_T^2)$,
\item to verify the self-consistency of the leading order QCD picture
of polarized SIDIS,
\item to check the validity of Lorentz invariance relations,
\item perform `global' phenomenological analysis by simultaneous
extraction of TMD DF's parameters from experimental data taking into
account the general positivity constraints~\cite{bacchetta} for TMD
DFs.
\end{itemize}
\bibliographystyle{aipproc}   % if natbib is available

\end{document}